\documentclass[final,times,5p]{elsarticle}
\usepackage{ctable}
\usepackage{ftnright}
\usepackage{graphicx}
\usepackage{xspace}
\usepackage{hhline}
\usepackage{amsmath}
\usepackage{amsfonts}
\usepackage{hyperref}

\newcommand{\ee}{\ensuremath{e^{+}e^{-}}\xspace}

\newcommand{\JP}{\ensuremath{J/\psi}\xspace}

\renewcommand{\epsilon}{\varepsilon}
\newcommand{\SKIP}[1]{}
\newcommand{\B}{\mathcal{B}}
\newcommand{\Gh}{\ensuremath{\tilde{\Gamma}_h}\xspace}

\begin{document}
%\begin{center}
%Revision of results on $\Upsilon(1S)$, $\Upsilon(2S)$, and
%$\Upsilon(3S)$ masses\\*[2ex]
%A.G. Shamov and O.L. Rezanova
%\end{center}
%\SKIP{
%--\begin{frontmatter}
\title{Revision of results on $\Upsilon(1S)$, $\Upsilon(2S)$, and $\Upsilon(3S)$
  masses}
%--------------
\author[binp]{A.G.~Shamov}
\author[binp,nsu]{O.L.~Rezanova}
 \address[binp]{Budker Institute of Nuclear Physics, 11, akademika
 Lavrentieva prospect,  Novosibirsk, 630090, Russia}
\address[nsu]{Novosibirsk State University, 1, Pirogova street, Novosibirsk, 630090, Russia}
%}

\begin{abstract} 
We have reconsidered the results on the masses of the narrow bottomonium states
$\Upsilon(1S)$--$\Upsilon(3S)$ obtained in 1982--1986 at CESR, DORIS and VEPP-4
colliders in order to fix shortcomings of the mass determination
procedures. For experiments at CESR and DORIS this includes the incorrect accounting
of the radiative corrections and usage of the electron mass value revised in 1986.
In analyses of all experiments the interference of the resonance production
and the nonresonant process was ignored. The corrected mass values for five
experiments are suggested. The corrections vary from 0.1 to 0.4~MeV.
The discrepancy between CESR and \mbox{VEPP-4} results on $\Upsilon(1S)$
mass has been reduced from 3.3 to 1.8 standard deviations.
\end{abstract}

\maketitle

\begin{table*}[t]
\caption{\label{tab:dM}Corrections to published mass values (keV).}
\begin{center}
\begin{tabular}{|l|r|r|r|r|r|}
\hline
 \multicolumn{1}{|c|}{$\Upsilon$--state} 
& \multicolumn{2}{c|}{$\Upsilon(1S)$} 
& \multicolumn{2}{|c|}{$\Upsilon(2S)$} 
& $\Upsilon(3S)$ \\
\hline
 \multicolumn{1}{|c|}{Collider} & CESR  &  VEPP-4  & DORIS &  VEPP-4 & VEPP-4 \\
\hline
 \multicolumn{1}{|c|}{$\sigma_W$ (MeV)} & 3.2  &  4.5  &  8.1 &  5.3 & 5.4 \\
\hline
Electron mass          & -81 &      & -86 &   &   \\
Radiactive corrections & -81 &      & -181 & & \\
Interference           & -71 & -112 & -168 & -105 & -130 \\
Resonance shape calculation      & +375 & & & & \\
\hline
Total  & +140 &     & -430 &    &    \\
\hline
Shift with correct fit  & +142 &  -112 & -435 & -105 & -130 \\
\hline
\end{tabular}
\end{center}
\end{table*}

\section*{Introduction}

The new experiment on the high precision  measurement
of the $\Upsilon$--meson mass has been planned at the VEPP-4M
collider~\cite{VEPP} with
the KEDR detector~\cite{KEDR:Det}.
The resonant depolarization method~\cite{REDE1,REDE2}
will be used for the beam energy determination.
At the moment the laser polarimeter is under development~\cite{LasPol}
and the test scan of the $\Upsilon(1S)$ has been performed~\cite{Pim}.
In this context it is important to overcome known drawbacks
in the analyses of the preceding experiments and correct its results.
 
The mass of $\Upsilon(1S)$ was measured by the MD-1 detector at
\mbox{VEPP-4}~\cite{mdU1S1982,mdU1S1986} and CUSB detector at
the collider CESR with the accuracy of about 0.1~MeV~\cite{cusbY1S}.
With lower accuracy of 0.4-0.5~MeV the mass of $\Upsilon(2S)$ was measured 
by ARGUS and Crystal Ball detectors at DORIS~\cite{dorisY2S} and
by MD-1~\cite{md1Y13}.  The mass of $\Upsilon(3S)$ was
determined with 0.5~MeV uncertainty by MD-1 only~\cite{md1Y13}.
In all these experiments the mass values were obtained by fitting the
inclusive hadronic cross section as function of the c.m. energy. The
beam energy was determined using resonant depolarization
method.

In 2000 the results
of the mass measurements at VEPP-4 were
corrected~\cite{Revisit} to the shift of the electron mass value
occurred in 1986~\cite{m_e}. The results from CESR and DORIS
stayed intact.

Another problem to solve is accounting of the radiative correction in
experiments~\cite{cusbY1S} and ~\cite{dorisY2S} according to the
work~\cite{JS} containing the mistake~\cite{KF}.
Despite to the existence of correct studies
of the narrow resonance production since 1975,
the incorrect resonance shape from Ref.~\cite{JS}
was employed for determination of leptonic
widths and masses of $\psi$-- and $\Upsilon$--states in many
experiments. Concerning leptonic widths
the problem was solved in Ref.~\cite{Alexander:1988em}, the corrected
values were included in PDG tables. 
However, the masses of $\Upsilon$ states~\cite{cusbY1S} and~\cite{dorisY2S}
were not corrected neither in
Ref.~\cite{Alexander:1988em} nor in
Ref.~\cite{Revisit} where 
radiative corrections were fixed up for $J/\psi$ mass measurement
by the OLYA detector in 1980~\cite{OLYA}.
The MD-1 experiments on masses of upsilon
states~\cite{mdU1S1982,mdU1S1986,md1Y13,mdU1S1992}
were performed with proper radiative
correction accounting.

Besides, there is a mistake in the calculation
of the resonance curve in Ref.~\cite{cusbY1S}, that will be discussed
in details below.

The common drawback of all measurements of $\psi$-- and $\Upsilon$-- state
masses mentioned above is ignoring
of the interference between resonant and
nonresonant contributions to the hadron production. First time it
was accounted in the $J/\psi$-- and $\psi(2S)$-- mass measurement in
the experiment~\cite{M2003} and was discussed in details in
Ref.~\cite{psi2S}. 

In all experiments under discussion except~\cite{cusbY1S}, the
dependence of the hadronic cross section on c.m. energy was not
published. In Refs.~\cite{Revisit,Alexander:1988em},
in order to correct the resonance leptonic width~\cite{Alexander:1988em}
and mass~\cite{Revisit}, the equidistant data
points were simulated using the published values of the resonance curve
parameters. Then two fits were performed with the correct fitting
function and that of the published paper. The variation of the
resonance parameter was added to its published value.
This method uses published values of parameters, biased by the
incorrect fit, and does not account for the specific layout of
energy points.

In contrast with that, in this work we obtained coordinates of data
points from the plots in electronic versions of publications using the
graphical editor and converted them to the physical quantities. Such
data were not absolutely reliable thus we shifted published
values
as described above. In Ref.~\cite{cusbY1S} the measured values of cross
section and the energy were published thus we could just refit the CUSB
data.

In the next sections we describe the necessary corrections, discuss
data published by CUSB~\cite{cusbY1S}, and then obtain corrected
mass values for three resonances from five experiments.

\section{Change of the electron mass value}

As it was mentioned above, the experiments cited did employ the
resonant depolarization method for the beam energy determination.
In this method the measured ratio of the spin precession frequency $\Omega$
and the revolution frequency $\omega$ gives the Lorentz factor
$\gamma$ according to the relation
\begin{equation}
\Omega/\omega = 1 + \gamma \cdot \mu^{\prime}/\mu_0,
\label{eqn:Omega}
\end{equation}
where $\mu^{\prime}$ and $\mu_0$ are anomalous and normal parts of the
electron magnetic moment~\cite{REDE2}. To find the beam energy
$E=\gamma\,m_e$, the value of the electron mass $m_e$ is required.
Before 1986 its accuracy was estimated to be 2.8 ppm~\cite{m_e_old}
which corresponds to 26~keV uncertainty in the mass of $\Upsilon(1S)$.
In 1986 the adjustment of fundamental physical constants~\cite{m_e}
led to shift of the electron mass value by -8.5~ppm with increase of
its accuracy to 0.8~ppm. The corresponding shifts of $\Upsilon(1S)$,
$\Upsilon(2S)$,  and $\Upsilon(3S)$ masses are -81, -86, and -88~keV
respectively~\cite{Revisit}. For experiments ~\cite{cusbY1S,dorisY2S}
that was not accounted yet.

\section{Radiative corrections}
Soon after the $J/\psi$--meson discovery a number of papers appeared on the
radiative corrections for a narrow resonance production in
$e^+e^-$--collisions.
First of them is probably Ref.~\cite{Azimov} which will be considered in the
next section. 
However, the most popular theoretical work used for $\psi$- and
$\Upsilon$- data analysis until 1985 was Ref.\cite{JS}. 
It was directly addressed to experimentalists and published in
``Nuclear Instruments and Methods''. The calculations were performed
in the approximation of zero resonance width. For the gaussian
collider energy spread distribution
\begin{equation}
G(W^\prime-W)= \frac{1}{\sqrt{2 \pi} \sigma_W} \,
 \exp \biggl ( -\frac{(W^\prime-W)^2}{2\sigma_W^2} \biggr )
\end{equation}
the narrow resonance cross section in a final state $f$
at the c.m. energy $W$ was obtained in the form
\begin{equation}\label{eq:JS}
\begin{split}
\sigma^{\text{R}\to f}(W) &=  \frac{6\pi^2}{M^2}\,
       \frac{\Gamma^{(0)}_{ee}\Gamma_f}{\Gamma}\,
 \biggl( G_r(W-M) + \delta\cdot G(W-M) \biggr), \\
G_r(x) & =  \left(\frac{2\sigma_W}{M}\right)^{\!\!\beta}
\frac{\Gamma(1+\beta)}{\sqrt{2 \pi} \sigma_W}
 \exp \left( -\frac{x^2}{4\sigma_W^2} \right)
 D_{-\beta}\left(-\frac{x}{\sigma_w}\right), \\
  \delta&=\frac{13}{12}\,\beta+
   \frac{2\alpha}{\pi}\left(\frac{\pi^2}{6}-\frac{17}{36}\right),
   \:\:
 \beta = \frac{4\alpha}{\pi} \left( \ln\frac{W}{m_e} -\frac{1}{2}\right),
\end{split}
\end{equation}
where $\alpha$ is the fine structure constant,
$m_e$ is the electron mass, 
$\Gamma()$ is the gamma-function and
 $D_{-\beta}$ is the Weber function of parabolic cylinder for
calculation of which the power series were specified. The
electron partial widths $\Gamma^{(0)}_{ee}$
corresponds to the lowest order of QED. The $\delta$
includes the vertex corrections and contribution of electron loops into the
vacuum polarization, other contributions are neglected. 

The $G_r$ function in \eqref{eq:JS} is so called ``radiative gaussian'',
which is a convolution of the collider energy spread with the probability
of the energy loss due to soft photon radiation in $e^+e^-$--collision.
It is known that the probability of the QED process which is not accompanied
by such emission is zero, therefore the second term in eq.~\eqref{eq:JS} is not
correct. There
must be $\delta\cdot G_r(W\!-\!M)$. The $G(x)$-function
unlike to $G_r(x)$ is symmetric, thus using of eq.~\eqref{eq:JS}
for data analysis increase the $\Upsilon(1S)$--mass by about 0.1~MeV at
the energy spread $\sigma_W\simeq5$~MeV, as was noted
in~\cite{mdU1S1986}.

\section{Interference effect}

The interference effects in the inclusive hadronic cross section in the
vicinity of a narrow resonance was considered already in Ref.~\cite{Azimov}.
With some up-today modifications the resonant and interference terms
in the soft photon approximation can be written as~\cite{psi2S}
\begin{equation}
\begin{split}
\sigma^{\text{r+i}}(W)&=\frac{12}{W^2}(1+\delta)
\left(
\frac{\Gamma_{ee}\tilde{\Gamma}_h}{\Gamma M}\,\mbox{Im} f(W)
-\frac{2\alpha \sqrt{R \Gamma_{ee}\tilde{\Gamma}_h}}{3W}
\lambda \, \mbox{Re} \frac{f^*(W)}{1-\Pi_0} \right)\!, \\
  \delta&=\frac{3}{4}\beta+
   \frac{\alpha}{\pi}\left(\frac{\pi^2}{3}-\frac{1}{2}\right)+
  \beta^2\left(\frac{37}{96}-\frac{\pi^2}{12}-
  \frac{1}{36}\ln\frac{W}{m_{\rm e}} \right)\!. \\
f(W) &=  \frac{\pi \beta}{\sin{\pi \beta}}\,
   \Bigg(\frac{W^2}{M^2-W^2-i M \Gamma} \Bigg)^{1-\beta}
\!\!\!\!\!.
\end{split}\label{eq:Azimov}
\end{equation}

Here $R$ is the hadron-to-muon cross section ratio out of the resonance peak,
 $\Pi_0$ is the vacuum polarization with the resonance
contribution excluded, $\Gamma_{ee}\!=\!\Gamma^{(0)}_{ee}/|1-\Pi_0|^2$ 
is the physical value of the electron width, $\tilde{\Gamma}_h$
is some effective value of the hadronic partial width, and
$\lambda$ is the parameter
introduced in
Ref.~\cite{Azimov} to characterize 
the strength of the interference
effects. Due to the interference of electromagnetic and strong decays
of the resonance, $\tilde{\Gamma}_h$ differs from the true hadronic
partial width $\Gamma_h$, but the value of \Gh is not of first importance
to the mass determination.  The problem is discussed in details in
Ref.~\cite{psi2S}. 
The following result was obtained for $\lambda$:
\begin{equation}
  \lambda = \sqrt{\frac{R \B_{ee}}{\B_h}} + \sqrt{\frac{1}{\B_h}}\,
        \sum\limits_m\!
    \sqrt{b_m \B^{(s)}_m\,}
          \left<\cos{\phi_m}\right>\,.
\label{eq:lambdaSum}
\end{equation}
The sum in Eq.~\eqref{eq:lambdaSum} is performed over all hadronic modes,
$\B_{ee}$ and $\B_h$ are the resonance decay probabilities
to \ee--pairs and hadrons, respectively,
$\B^{(s)}_m\!=\!\Gamma^{(s)}_m/\Gamma$
is related to the strong contribution to the decay mode $m$,
process, $\phi_m$ is its phase relative to the electromagnetic
contribution and $b_m\!=\!R_m/R$ is the branching fraction of
the corresponding continuum. The angle brackets denote averaging
over the decay products phase space.

Following Ref.~\cite{M15}, we assumed that the relative phases of the strong
and electromagnetic amplitudes in different decay modes are not
correlated, thus the second term of Eq.~\eqref{eq:lambdaSum} can be
neglected compared to the first one, which is about 0.31, 0.27 and 0.29
for $\Upsilon(1S)$, $\Upsilon(2S)$ and $\Upsilon(3S)$, respectively.
The same value of $\lambda$ follows from naive parton model in which
$q\bar{q}$, $ggg$ and $gg\gamma$ decay modes are considered.
For $J/\psi$-meson the measurement gave $\lambda_{\JP} = 0.45\pm0.07\pm0.04$
at the expected value of 0.39~\cite{M15}. Scaling this result 
to $\Upsilon(nS)$, one obtains the mass uncertainty estimate
of about $4.10^{-3}\sigma_W$.
The correction grows with the value of the energy spread,
see Table~\ref{tab:dM} above.

\begin{figure}[!tb]
\vspace*{-5mm}
%\hspace*{10mm}
\centering
\includegraphics[width=0.95\columnwidth]{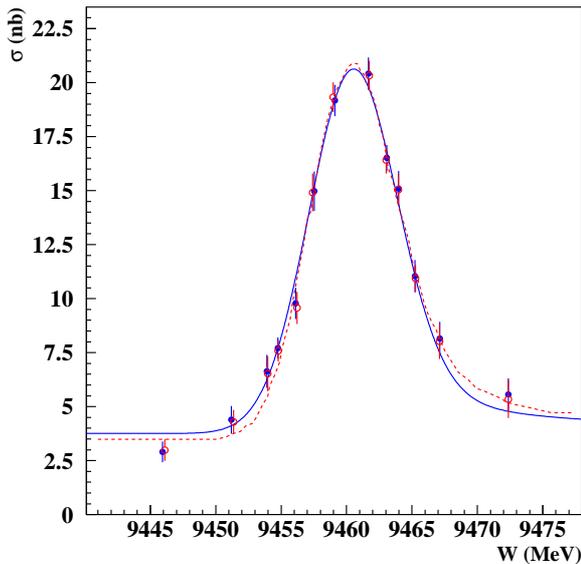}
\vspace*{-6mm}
\caption{\label{fig:CUSB}(Color online) The profile of the $\Upsilon(1S)$ 
as measured with the CUSB detector~\cite{cusbY1S}.
The red open circles and the dashed curve are digitized from Fig.~10 of
the paper. The blue closed circles correspond to the data points from the
published table. The blue curve shows our fit with Eq.~\eqref{eq:JS}.}
\end{figure}

\section{Reanalysis of CUSB data}

During the $\Upsilon(1S)$ mass measurement at the CESR collider
with the CUSB detector, 22 runs were recorded, which were jointed
in 11 data points
for the fit. The beam energy during runs were
determined using
the value of bend magnetic field measured with NMR according to linear
relation $E=A(B\!-\!B_0)\!+\!C$, where $B$ is a measured NMR value 
and $B_0$ is some reference one.
The constants $A$ and $C$ were obtained by fitting of
10 measurements of the beam energy with the resonance depolarization method.
 
The data point number, the NMR value, the number of hadronic events,
the integrated luminosity and the cross section for each run were presented
at Table~I of Ref.~\cite{cusbY1S}.

In Figure~1 the data points calculated by us using Table~I and published
values of $A$, $C$ and $B_0$ are compared with these extracted
from Fig.~10 of Ref.~\cite{cusbY1S} using the GIMP graphical editor.
Both energies and cross sections of the points agree within the
accuracy achieved with the editor.
However, the curves of the fits performed using the same formulae
and with the same value of the electron mass differ. We have checked
our calculation comparing the results obtained in the zero-width
approximation using two different implementations of the Weber function
and those obtained using numerical convolution of Eq.\eqref{eq:Azimov}
with the gaussian energy spread.
The three our results agree with each other thus we conclude that
the mass value published in Ref.~\cite{cusbY1S} is not fully correct
and should be shifted by +0.375~MeV.  

There is a question to Table~I concerning assignment of the run 14 to the
point 8, its NMR value is closer to those of the point 9. 
This might be a misprint. Blue closed points in Figure 1 are shown for
the proper run-to-point assignment. The corresponding change of the mass
value is negligible.

\section{Values of corrections to $\Upsilon(1S)$--$\Upsilon(3S)$ masses}

The corrections to mass values obtained in five experiments due to
effects considered are presented in Table 1. The sum of corrections
calculated separately is
in a good agreement with the shift obtained
with the correct fit. The uncertainty of reanalysis due to accuracy
of the vacuum polarization data is 1--2 keV. The similar error is connected
to digitization of journal plots.  They do not reduce the accuracy of
experimental data.

\section*{Conclusion}

The results of five experiments on the precision measurements of masses
of narrow $\Upsilon$ states were reanalyzed to remove substantial drawbacks of
original analyses. The following results were obtained:\
\begin{eqnarray}
\text{M}_{\Upsilon(1S)}= &9460.40 \pm 0.09&\!\!\!\!\pm\: 0.04\:\:\text{MeV}
\:\text{(MD-1 \cite{mdU1S1992})}.\nonumber\\
\text{M}_{\Upsilon(1S)}= &9460.11 \pm 0.11&\!\!\!\!\pm\: 0.07\:\:\text{MeV}
\:\text{(CUSB \cite{cusbY1S})}.\nonumber
\end{eqnarray}

\vspace*{-5ex}
\begin{eqnarray}
\text{M}_{\Upsilon(2S)}= &10023.4 \pm 0.5&\!\!\text{MeV}\:\text{(MD-1 \cite{md1Y13})}.\nonumber\\
\text{M}_{\Upsilon(2S)}= &10022.7 \pm 0.4&\!\!\text{MeV}\:\text{(ARGUS+CB \cite{dorisY2S})
}.\nonumber\\*[0.5ex]
\text{M}_{\Upsilon(3S)}= &10355.1 \pm 0.5&\!\!\text{MeV}\:\text{(MD-1 \cite{md1Y13})}. \nonumber
\end{eqnarray}
The discrepancy between MD-1 and CUSB results on the 
$\Upsilon(1S)$ mass has been reduced
from 3.3 to 1.8 standard deviations.
The mean value of two experiments, calculated according the PDG rules
with the scale factor of 1.8 is
\begin{equation}\nonumber
\text{M}_{\Upsilon(1S)} = 9460.29 \pm 0.15\:\text{MeV}.
\end{equation}
The uncertainty is reduced from 0.33 to 0.15~MeV.

\vspace*{1.5ex}
We appeal the Particle Data Group to accept these mass values as it was
done with the leptonic widths recalculated in
Ref.~\cite{Alexander:1988em}.

\end{document}